\documentclass[10 pt,letterpaper]{article}
\usepackage{opex3}
\usepackage{float}

\begin{document}

\title{Effect of Speckle on APSCI method and Mueller Imaging}

\author{Debajyoti Upadhyay\textsuperscript{1}, Michael Richert\textsuperscript{2}, Eric Lacot\textsuperscript{3}, Antonello De Martino\textsuperscript{2} and Xavier Orlik\textsuperscript{1*}}

\address{\textsuperscript{1}ONERA, Theoretical and Applied Optics Department, 31055 Toulouse, France\\
\textsuperscript{2}LPICM, Ecole Polytechnique, CNRS, 91128 Palaiseau, France\\
\textsuperscript{3}Laboratoire de Spectrom\'etrie Physique (UMR 5588), Universit\'e Joseph Fourier, CNRS, 38402 St Martin d'H\`eres, France}
\email{\textsuperscript{*}xavier.orlik@onera.fr} 




\begin{abstract} The principle of the polarimetric imaging method called APSCI (\textbf{\textit{A}}dapted \textbf{\textit{P}}olarization \textbf{\textit{S}}tate \textbf{\textit{C}}ontrast \textbf{\textit{I}}maging) is to maximize the polarimetric contrast between an object and its background using specific polarization states of illumination and detection. We perform here a comparative study of the APSCI method with existing Classical Mueller Imaging(CMI) associated with polar decomposition in the presence of fully and partially polarized circular Gaussian speckle. The results show a noticeable increase of the Bhattacharyya distance used as our contrast parameter for the APSCI method, especially when the object and background exhibit several polarimetric properties simultaneously.
\end{abstract}

\ocis{(120.5410) Polarimetry; (260.5430) Polarization; (110.5405) Polarimetric imaging; (110.2970) Image detection systems} 


\section{Introduction}

The polarimetric imaging method \textbf{\textit{APSCI}}\cite{RICHERT2009} has been shown to reach beyond the limit of contrast achievable from the Classical Mueller Imaging (CMI) with polar decomposition \cite{LU1996}. The process utilises a selective polarimetric excitation of the scene in order to provoke a scattering from the object and background characterized by Stokes vectors as far as possible in the Poincar\'e Sphere \cite{POINCARE1892}. Then along with an optimal polarimetric detection method specifically adapted to each situation, it has been demonstrated that the contrast between an object and its background could be increased to a higher order of magnitude with respect to the contrast from CMI with polar decomposition \cite{RICHERT2009}.

We propose here to study the performance of the APSCI method taking into account the shot noise of the detector and the speckle noise in the case of a monochromatic illumination giving rise to an additional circular Gaussian speckle noise, where partial depolarization may occur. Moreover, we consider the numerical propagation of
errors in the calculation of the polarimetric data from the acquired raw data. For various situations, where the scene exhibits different polarimetric properties such as dichroism, birefringence or depolarization, we perform a comparative study of contrast level, which is quantified by the Bhattacharyya distance  \cite{BHATTACHARYYA1943} calculated from the significant parameters of the CMI, the polar decomposition and the APSCI method.

\section{Brief review of the APSCI method}  
\label{APSCIREV}

Let us assume a scene with a homogeneous circular object surrounded by a homogeneous background. Then, the scene can be modeled into two mutually exclusive regions $\mathcal{O}$ and $\mathcal{B}$ having polarimetric properties characterized by their Mueller matrices $\mathbf{M_O}$  and $\mathbf{M}_B$, respectively for the object and the background. As the scene is considered to be a priori unknown, we need an initial estimation of Mueller matrices of the object $\mathbf{\widetilde{M}_O}$ and of the background $\mathbf{\widetilde{M}_B}$ by CMI before implementing APSCI method. During the Mueller imaging process, we consider that each pixel of the detector indexed by $(u,v)$ receives an intensity $\mathbf{I}(u,v)$  perturbed by a Poisson distribution in order to take into account the shot noise. The Mueller matrix $\mathbf{\widetilde{M}}(u,v)$ at each pixel is then calculated from the noisy detected intensities $\mathbf{\widetilde{I}}(u,v)$.

Let us assume a totally polarized Stokes vector $\vec{S}$ is used to illuminate the scene after CMI. The estimations of the Stokes vectors of the field scattered by the object $\widetilde{\vec{S}}_O$ and background $\widetilde{\vec{S}}_B$ can be expressed as :
\begin{equation} \label{S_scene}
	\textbf{$\widetilde{\vec{S}}_O$} =
	 \left[S_{O_0},\ S_{O_1},\ S_{O_2},\ S_{O_3}\right]^T	= \mathbf{\widetilde{M}_O}\vec{S}\ , \quad
	 \textbf{$\widetilde{\vec{S}}_B$} =
	 \left[S_{B_0},\ S_{B_1},\ S_{B_2},\ S_{B_3}\right]^T	= \mathbf{\widetilde{M}_B}\vec{S}\ 
\end{equation}

We define the measure of separation of $\widetilde{\vec{S}}_O$ and $\widetilde{\vec{S}}_B$ in the Poincar\'e sphere by the Euclidean distance $D$ between their last three parameters. Then, we determine numerically using a simplex search algorithm the specific incident Stokes vector $\vec{S}^{in}$ that  maximizes this Euclidean distance. It is worthy to emphasize that the Stokes vectors $\widetilde{\vec{S}}_O$ and $\widetilde{\vec{S}}_B$ are not normalized and can exhibit different rate of depolarization. As a consequence, the maximization of the Euclidean distance mentioned above takes into account two physical entities : the polarization state and the intensity of the polarized part of the scattered field. 

Finally, we utilize a Two Channel Imaging (TCI) system that projects the scattered field resulting from the selective excitation $\vec{S}^{in}$, into 2 states of polarization $\vec{S}^{out 1}$ and $\vec{S}^{out 2}$, that are defined to maximize respectively ($\widetilde{I}_O-\widetilde{I}_B$) and ($\widetilde{I}_B-\widetilde{I}_O$), where $\widetilde{I}_O$ and $\widetilde{I}_B$  are the evaluations of the mean intensity detected respectively from the object and background scattering. From simple calculation it can be shown that:

\begin{equation}
\vec{S}^{out 1} = \left[1,\ \Delta \vec{S}^T\ / \left\| \Delta \vec{S} \right\| \right]^T\ ,\quad \vec{S}^{out 2} = \left[1,\ - \Delta \vec{S}^T\ / \left\| \Delta \vec{S} \right\| \right]^T
\end{equation} 

where
\begin{equation}
	\Delta \vec{S} = \left[\widetilde{S}_{O_1}^{out}-\widetilde{S}_{B_1}^{out},\ \widetilde{S}_{O_2}^{out}-\widetilde{S}_{B_2}^{out},\ \widetilde{S}_{O_3}^{out}-\widetilde{S}_{B_3}^{out} \right]^T
,\end{equation}

with
\begin{equation} 
\left[\widetilde{S}_{O_0}^{out},\widetilde{S}_{O_1}^{out},\widetilde{S}_{O_2}^{out},\widetilde{S}_{O_3}^{out} \right]^T = \widetilde{\vec{S}}_{O}^{out} = \mathbf{\widetilde{M}_O}\vec{S}^{in}\ , \quad
\left[\widetilde{S}_{B_0}^{out},\widetilde{S}_{B_1}^{out},\widetilde{S}_{B_2}^{out},\widetilde{S}_{B_3}^{out} \right]^T = \widetilde{\vec{S}}_{B}^{out} = \mathbf{\widetilde{M}_B}\vec{S}^{in}\
.\end{equation}
 
The APSCI parameter is then defined for each pixel of the detector indexed by the coordinates $(u,v)$ as :

\begin{equation} \label{APSCIeq}
APSCI(u,v) = \frac{I_1(u,v)-I_2(u,v)}{I_1(u,v)+I_2(u,v)},
\end{equation}
where $I_1(u,v)$ and $I_2(u,v)$ are the detected intensity after projection respectively on the 2 states of polarization $\vec{S}^{out 1}$ and $\vec{S}^{out 2}$.\\
In this study, we use the Bhattacharyya distance as a contrast parameter for each physical quantity under investigation that can be, for comparison purposes, either the APSCI parameter as defined above, either the more pertinent parameters extracted from the polar decomposition of the Mueller matrices of the object and background. A more detailed discussion of the APSCI method is proposed in \cite{RICHERT2009}.
\\

\section{Characteristics of the Speckle noise}
We have chosen to study an unfavourable situation of imaging regarding both the speckle grain size and its contrast.
Thus, we assume a speckle grain with a size similar to that of the pixel of the detector. On the experimental point of view, this situation corresponds to a contrast that is not decreased by the integration of several grains into a single pixel.\\
Moreover, we choose to study the effect of a completely polarized and developed circular Gaussian speckle because it exhibits a strong contrast and so is susceptible to decrease the performance of the APSCI method. As will be pointed out later in this article, biological applications of the APSCI method seem very promising. So, in order to take into account some possible movement of the object, we consider a dynamical speckle: each intensity acquisition is then submitted to a different speckle pattern.\\
The effect of a partially polarized speckle is also of interest regarding the APSCI method because it combines 2 antagonist effects : a decrease of the speckle contrast that increases the Bhattacharyya distance of the APSCI parameter and a lower amount of polarized light usable by the APSCI method for the optimization that, on the contrary, is expected to decrease signal to noise ratio and hence this distance.
In our simulations, the speckle is taken into account by a modulation of intensity at the image plane that is considered independent of the state of polarization scattered by the object and background. This modulation of intensity is performed according to the probability density function of intensity $p_\textsl{I}(I)$ of a completely developed circular Gaussian speckle that depends on the degree of polarization $\emph{\textsl{P}}$ \cite{GOODMAN2007}:

\begin{equation} \label{partpol}
p_\textsl{I}(I) = \frac{1}{\emph{\textsl{P}}\overline{I}}\left[exp\left(-\frac{2}{1+\emph{\textsl{P}}}\frac{I}{\overline{I}}\right)-exp\left(-\frac{2}{1-\emph{\textsl{P}}}\frac{I}{\overline{I}}\right)\right]
\end{equation}

where $\overline{I}$ is the average intensity.

\section{Results and analysis}

\begin{figure}[H]
	\centering
		\includegraphics[width=16cm,height=16cm]{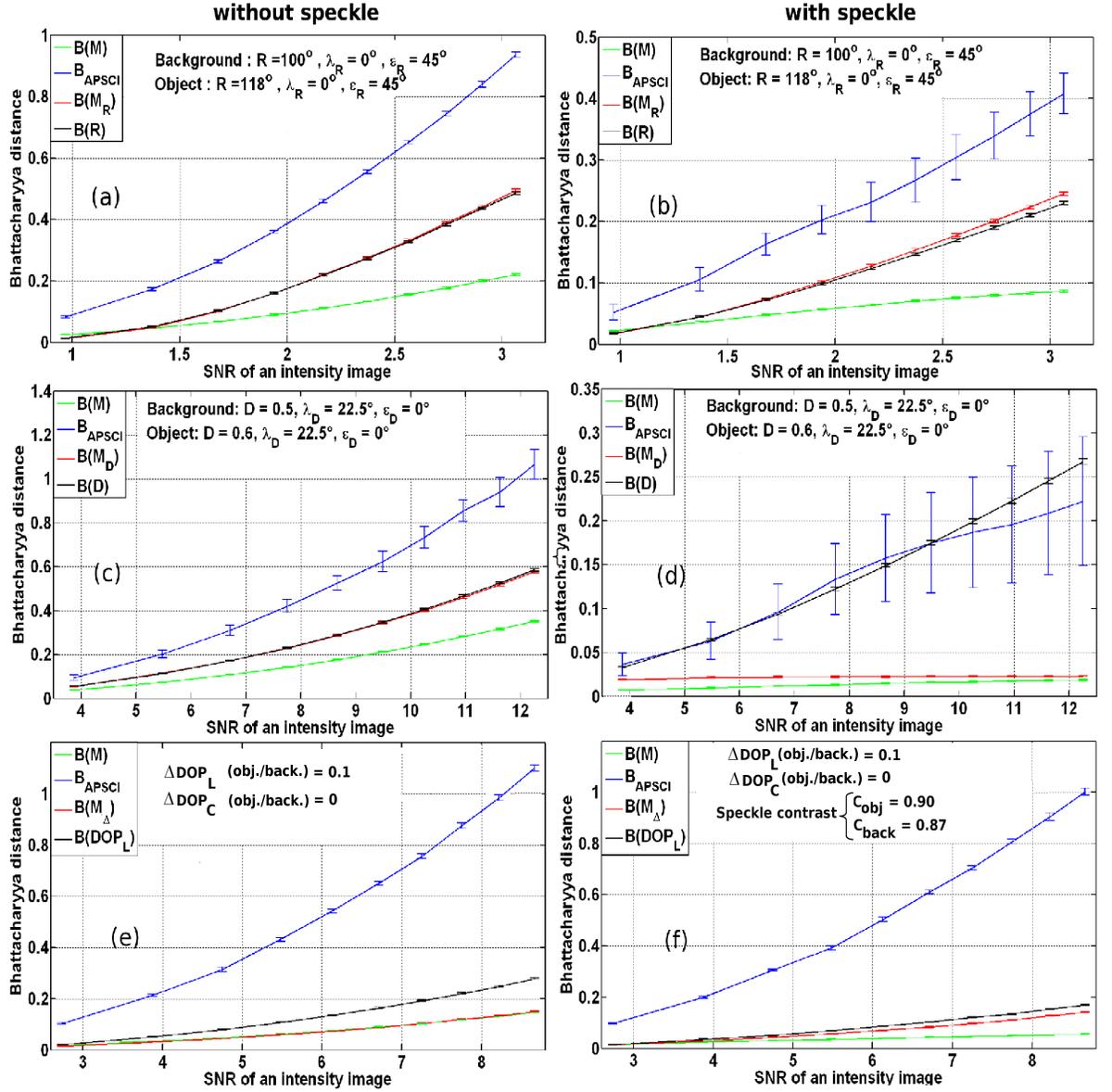}
	\caption{Bhattacharyya distances obtained from different SNR levels without (case a, c, e) and with (case b, d, f) speckle noise. The object and background have a difference of 10\% only in scalar birefringence, scalar dichroism and linear degree of polarization respectively for the cases a \& b, c \& d and e \& f. In case a \& b, (R, $\lambda_R$, $\epsilon_R$) represent respectively the scalar birefringence, azimuth and ellipticity of the birefringence vector $\vec{\mathbf{R}}$. Similarly for case c \& d , (D, $\lambda_D$, $\epsilon_D$) represent respectively the scalar dichroism, azimuth and ellipticity of the dichroism vector $\vec{\mathbf{D}}$. For the case e \& f, eigen axes of depolarization of the object and background are assumed aligned and $\Delta DOP_L$ and $\Delta DOP_C$ represent respectively the difference between the object and background, in their ability to depolarize linear and circular polarized light.}
	\label{fig:test1}
\end{figure}

\begin{figure}[H]
	\centering
		\includegraphics[width=8cm,height=5cm]{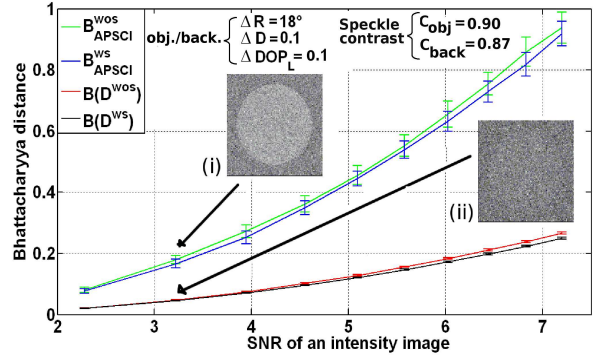}
	\caption{Comparison of Bhattacharyya distances vs. SNR curves for the best performing parameter of CMI (in this case the scalar dichroism) vs APSCI parameter with (ws) and without speckle noise (wos). The scene is composed of an object and a background exhibiting 10\% difference in scalar birefringence R, scalar dichroism D and in the linear degree of polarisation $DOP_L$. At the same SNR level of 3.2, the embedded images (i) and (ii) are obtained respectively using the APSCI parameter and the best parameter of the CMI.}
	\label{fig:test2}
\end{figure}

We have chosen to study in Fig. \ref{fig:test1} the effect of a completely developed circular Gaussian speckle on three different situations where the object and background are defined to have a difference of 10\% in one polarimetric property : the cases (a) and (b) exhibit this difference in the scalar birefringence, the cases (c) and (d) in the scalar dichroism and the cases (e) and (f) in the degree of linear polarization. The situations (a), (c) and (e)  consider only the shot noise whereas (b), (d) and (f) take into account an additional speckle noise. For each of these situations, we calculate between the object and background region, the Bhattacharyya distance of the APSCI parameter and of the other pertinent parameters extracted from the polar decomposition. For comparison purposes, the Bhattacharyya distances are plotted versus the signal to noise ratio (SNR) for a same number of intensity acquisition. We would like to point out that, due to their different dichroism, the energy scattered by the object and background and focalized by imaging elements towards the detector can be different, and hence their corresponding classical SNR's. Thus, we choose to define here a global SNR by considering the shot noise generated by the amount of energy received by the detector without the use of any polarizer and after the back-scattering on a virtual perfectly lambertian and non absorbing object, whose size and position are similar to that of the scene under investigation.
\\
In Fig. \ref{fig:test1} $B(M)$ is defined to be the selected element between $\mathbf{\widetilde{M}_O}$ and $\mathbf{\widetilde{M}_B}$  that provides the best Bhattacharyya distance over the 16 possible elements. In a similar way, $B(M_R)$, $B(M_D)$ and $B(M_\Delta)$ represent respectively the best Bhattacharyya distance obtained from the selected element of the birefringence, the dichroism and the depolarization matrices extracted from $\mathbf{\widetilde{M}_O}$ and $\mathbf{\widetilde{M}_B}$ using the forward polar decomposition.
\\
The Bhattacharyya distances corresponding to scenes exhibiting a difference of scalar birefringence,  scalar dichroism and in their ability to depolarize linear polarized light are plotted respectively as $B(R)$, $B(D)$ and $B(DOP_L)$. Finally, $B_{APSCI}$ represents the Bhattacharyya distance of the APSCI parameter as defined in section \ref{APSCIREV}.
\\
Considering the situations (a), (c) and (e) that take into account only the shot noise, we observe that from a SNR threshold, the polar decomposition that isolates the property of interest (red and black curves) brings always better Bhattacharyya distances than $B(M)$ (green curve) selected from the raw data of $\mathbf{\widetilde{M}_O}$ and $\mathbf{\widetilde{M}_B}$. Below this threshold, the noise introduced by this decomposition worsen the situation (as can be seen in case (a)) because $\mathbf{\widetilde{M}_O}$ and $\mathbf{\widetilde{M}_B}$ are insufficiently determined. Secondly, we observe that the parameter $B_{APSCI}$ (blue curve) exhibits the highest Bhattacharyya distances for all the SNR studied in (a) (c) and (e). However, for the case (c), we notice that it exhibits also higher uncertainty bars associated to lower mean values of Bhattacharyya distances compared to cases (a) and (e). This lower performance of the APSCI method in the case of dichroism is coming from 2 phenomena : the absorption of energy due to the dichroism effect and the cartesian distance between the matrices of the object and background defined here as the square root of the sum of the square of the element-wise differences. Indeed, as previously discussed in \cite{RICHERT2009}, a $10\%$ difference in one polarimetric property between the object and background gives rise to various cartesian distances in function of the scene studied. For cases (a), (c) and (e), the cartesian distances are respectively : 0.44, 0.09 and 0.14. The lowest value corresponds to the dichroism case and explains the lower performance of the APSCI method in that case.
\\
When adding the speckle noise, we observe in (b) compared to (a), a strong degradation of all the Bhattacharyya distances under study. $APSCI$ still remains the more pertinent parameter to distinguish the object from the background even if its standard deviation noticeably increases due to the presence of speckle. The effect of the same speckle noise on situation (c) is plotted on (d). We observe that $B(M_D)$ and $B(M)$ fall to very low values even for high SNR and as a consequence become unusable for imaging. As a result, from the raw data of the Mueller matrices $\mathbf{\widetilde{M}_O}$ and $\mathbf{\widetilde{M}_B}$ associated to the polar decomposition, only $B(D)$ reaches an order of magnitude similar to $B_{APSCI}$. Moreover, we notice that the standard deviation of $B_{APSCI}$ has considerably increased due to the speckle noise. After a deeper analysis, we have observed that $\mathbf{\widetilde{M}_O}$ and $\mathbf{\widetilde{M}_B}$ are particularly poorly estimated for the case of dichroism (for the 2 reasons mentioned above) and that the addition of speckle noise worsen noticeably this situation. As a consequence, selective states of excitation $S^{in}$  spread near all over the Poincar\'e sphere, showing only a weak increase of density of probability in the theoretical optimum region.
\\
We consider now on Fig.\ref{fig:test1} (e) and (f) the effect of a partially polarized speckle noise with degrees of polarization being respectively $P_{obj}=$0.78  and $P_{back}=$0.71 for the object and background that exhibit a difference of 10$\%$ in their ability to depolarize linear polarized light. We observe only a weak decrease of all the Bhattacharyya distances due to the fact that the speckle, only partially polarized, exhibits a lower contrast ($C_{obj}=0.90$ and $C_{back}=0.87$) than in previous situations. Moreover, the APSCI parameter gives rise to Bhattacharyya distances much higher than using the CMI alone or associated with the polar decomposition.
\\
In all the previous situations, we have studied scenes that exhibit a difference between the object and background in only one polarimetric property. However, in such pure cases, due to the numerical propagation of errors, the interest of using Mueller Imaging can be inappropriate compared to simpler methods such as ellipsometry \cite{DRUDE1889}\cite{JUNG2008} or polarization difference imaging methods \cite{ROWE1995}\cite{TYO1996}. However, Mueller Imaging can be of great interest in the case of scenes exhibiting several polarimetric properties at the same time. 
\\
Thus, in order to examine the performance of APSCI method in such case, we consider a more complex scene where the object and background have 10\% difference in scalar birefringence, scalar dichroism and in the linear degree of polarization simultaneously. 
In Fig. \ref{fig:test2}, we show the Bhattacharyya distances of the APSCI parameter compared to the best Bhattacharyya distances obtained from the CMI associated to the polar decomposition that is, in this new situation, the Bhattacharyya distance corresponding to scalar dichroism. We observe that both parameters show only a weak decrease of performance (around 5\%) due to the speckle noise because it is only partially polarized and exhibits a contrast significantly inferior to 1. Secondly, we see clearly that the Bhattacharyya distances of the APSCI parameter exhibits much higher values than the ones of the scalar dichroism. A visual comparison for a SNR of 3.2 is proposed in Fig. \ref{fig:test2} where the object can clearly be seen only using the APSCI parameter because of having 3.8 times higher Bhattacharyya distances compared to the one of the scalar dichroism.
\\ 
We would like to point out that the APSCI parameter of this complex scene exhibits better Bhattacharyya distances than the pure cases of dichroism and depolarization studied separately in Fig. \ref{fig:test1}.
\\ 
However, inspite of the same amount of birefringence in the situations of Fig. \ref{fig:test2} and Fig. \ref{fig:test1}(a) and that there are additional properties (dichroism and depolarization) in the mixed case which could help us to differentiate the object from the background, the Bhattacharyya distances obtained in the pure case of birefringence are higher than in the mixed case for a given SNR, even if we correct the SNR value by taking into account absorption that occurs in the latter case. This can be explained by the cartesian distances between the object and background matrices which are respectively 0.44 and 0.13 for the pure birefringent and mixed case. It can appear surprising that adding some polarimetric differences between the object and background can reduce the cartesian distance between their Mueller matrices and so our ability to distinguish them. 

However, we have to keep in mind that most of the elements of a Mueller matrix describe several properties simultaneously and that from a qualitative point of view, the effects of birefringence and dichroism can produce counter-effects that will decrease the maximum distance achievable between the Stokes vector scattered by the object and background. Such observation is true for all kind of polarimetric measurements, however, as the APSCI method doesn't consider each polarimetric property separately, and rather takes into account the whole Mueller matrices, it is expected to give always the best polarimetric contrast achievable for a 2 channel imaging system.

\section{Conclusion}
We have studied in various situations the effect of a completely or partially polarized and fully developed circular Gaussian speckle in presence of shot noise, on the APSCI method compared to the Classical Mueller Imaging associated to the polar decomposition. In spite of the additional high contrast speckle noise, the use of selective polarization states of illumination and detection in the APSCI method improves noticeably the polarimetric contrast between an object and its background with respect to the one achievable from the Classical Mueller imaging, even when pertinent polarimetric data are extracted by the polar decomposition. Moreover, as APSCI optimizes the polarimetric contrast combining dichroism, birefringence and depolarization properties simultaneously, it exhibits remarkably high performance compared to Classical Mueller Imaging when the object and background exhibit multiple polarimetric behaviour. This previous remark makes this technique very promising for medical applications as biological tissues often exhibit several polarimetric properties simultaneously in presence of dynamical speckle noise. Especially, it can represent a powerful and non invasive technique for accurate detection of displasic areas in case of tumour ablation. 

\section{Acknowledgement}
We would like to thank the region Midi-Pyr\'en\'ee for providing us financial support for this work.
\end{document}